%% file: WBF.tex
\documentclass[notoc,paper,12pt]{JHEP3}
   
\usepackage{amsmath}
\usepackage{graphicx}
\author{C.-P.~Buszello\\ Cavendish Laboratory, University of Cambridge, \\
  E-mail: \email{buszello@hep.phy.cam.ac.uk}}

\author{P.~Marquard\\ 
  Institut f\"ur Theoretische Teilchenphysik, Universit\"at Karlsruhe, \\
  E-mail: \email{Marquard@particle.uni-karlsruhe.de}}

\preprint{TTP06-10\\
SFB/CPP-06-11\\
CAVENDISH-HEP-06-10\\
hep-ph/0603209\\}

\title{Determination of Spin and CP of the Higgs Boson from WBF}

\abstract{We explore the possibilities to determine the spin/CP
  properties of the Higgs boson at the LHC. To cover the mass region
  below the ZZ threshold we make use of the properties of the production in Weak Boson Fusion (WBF)
  and the decay chain H $\rightarrow$ W$^+$W$^-$ $\rightarrow$ $\ell^+\nu \ell^-\nu$. 
  In particular, we study the angular correlations of the 
  forward jets and the distribution of the invariant mass of the lepton pair for
  different hypothetical Higgs like particles.}
\begin{document}

\input{introduction}

\input{chapter3}

\section{Conclusions}
We have implemented a matrix element generator capable of simulating the
generation of Higgs bosons via WBF and its decay to W pairs with SM and
non SM spin and CP quantum numbers (namely $0^{+}, 0^{-}, 1^+, 1^-$). We
have demonstrated that the different couplings lead to distinct
distributions of the opening angle of the tag jets and the invariant
mass of the leptons from the leptonic decay of the W pairs. The latter
distributions are very sensitive to the lepton cuts used to isolate the
signal.  This leads to the necessity of optimizing these cuts carefully
when performing the measurement using real data, depending on the
background level observed in the experiment. The tag jet distributions
appear much more robust against the cuts. They provide very promising
prospects to confirm the spin/CP state of an SM Higgs with a mass
between 130 and 180 GeV using an integrated luminosity of 30fb$^{-1}$.

\end{document}

%% file: introduction.tex
\section{Introduction}
The Standard Model (SM) is well established and in agreement with all present
collider data. The only part of the model, not explored so far, is the
Higgs sector. Because this sector plays a distinguished role in the
theory, being responsible for the masses and mixings of all particles,
the search for the Higgs boson is one of the highest priorities at the
LHC. Within the Standard Model all properties of the Higgs boson are
fixed when its mass is known. From indirect limits the Higgs boson is
expected to have a mass in the range 114.4~GeV\ $< m_H <$ 246~GeV\ 
(95\% C.L.) \cite{Eidelman:2004wy}. When the Higgs mass is above the
$Z$ pair threshold, it decays with a large branching fraction into $Z$
bosons, that can be discovered in the ``golden"
$\ell^+\ell^+\ell^-\ell^-$ decay mode. As long as $m_H \gtrsim$
130~GeV\ the decay into four leptons can still be used. Within the
region 155~GeV\ $< m_H <$ 170~GeV\ the $ZZ^*$ branching fraction
goes through a minimum, while the $WW^*$ decay mode opens up.  In this
mass range the $gg \to H \to WW \to \ell \nu \ \ell \nu$
\cite{Dittmar:1996ss} and the recently established vector boson fusion mode
$qq \to qq H \to qq WW \to qq \ \ell \nu \ \ell \nu$
\cite{Rainwater:1999sd,Asai:2004ws,cms-higgs} have the largest discovery
potential. 

In recent papers \cite{Buszello:2002uu, Choi:2002jk} we discussed 
the possible determination of the spin/CP
properties of the Higgs boson. To analyse these properties we introduced
hypothetical couplings to the Z-bosons corresponding to a Higgs-like
particle with non Standard Model spin and CP. To distinguish different
spin/CP eigenstates we considered the decay chain $H\to ZZ \to
\ell^+\ell^-\ell^+\ell^-$ and analysed the angular correlations between the leptons.
In \cite{Buszello:2002uu} we discuss pure spin/CP states up to spin 1,
\cite{Choi:2002jk} also considers spin 2 particles performing a similar
analysis. In addition we also consider mixed CP states for spin 0
particles in \cite{Buszello:2004be}. These analyses showed that the
spin/CP properties can easily be determined for a Higgs mass above the
threshold for Z pair production. Below the threshold it is more
challenging to obtain a significant separation.

To cover this mass region we now investigate Higgs
production via WBF and the subsequent decay chain $H\to WW\to \ell\nu\,\ell\nu$. Promising observables in this case are the angle
between the two forward jets and the invariant mass of the charged leptons.

A similar analysis using only the angle between the tag jets has already
been performed in \cite{Rainwater:1999sd}. They considered additional
dimension six operators to couple a spin 0 Higgs boson to vector bosons
including a CP odd and a CP even coupling not present in the Standard Model.

The NLO corrections to Higgs production via WBF are given in
\cite{Figy:2003nv}. They do not significantly
change the shape of the distributions and therefore we limit ourselves
to the leading order approximation.

The paper is organized as follows. In chapter 2 we briefly review the
model used for the non Standard Model couplings, in chapter 3 we
discuss the angular distributions of the tag jets and the invariant
mass of the lepton pair and finally we conclude.

\section{Model}
We use the same parametrisation as introduced in
\cite{Buszello:2002uu} and only repeat it here for completeness.

The most general coupling of a (pseudo) scalar Higgs boson with mass $M_h$
to two on-shell vector bosons is of the following form:
\begin{equation}
        {\cal L}_{scalar}=
           \mathbf{X} \delta_{\mu \nu}+
\mathbf{Y} k_{\mu} k_{\nu}/M_h^2 +i \mathbf{P} \epsilon_{\mu \nu p_{Z}
           q_{Z}}/M_h^2  .
\label{scalar}
\end{equation}
Here the momentum of the first boson is $p_Z^{\mu}$, 
that of the second boson is $q_Z^{\nu}$.
The momentum of the Higgs boson is $k$ and  $\epsilon_{\mu\nu \rho\sigma}$ is the total antisymmetric tensor with $\epsilon_{1234} = i$.
Within the Standard Model one has $\mathbf{X}=1$, $\mathbf{Y}=\mathbf{P}=0$.
For a pure pseudoscalar particle one has $\mathbf{P} \not= 0, \mathbf{X}=\mathbf{Y}=0$.
If both $\mathbf{P}$ and one of the other interactions are present,
one cannot assign a definite parity to the Higgs boson.

A similar formula  for a (pseudo) vector with momentum
$k_{\rho}$ reads:
\begin{equation}
        {\cal L}_{vector}=
         \mathbf{X} (\delta_{\rho \mu} p_Z^{\nu}+\delta_{\rho \nu} q_Z^{\mu})
                  +\mathbf{P} (i \epsilon_{\mu \nu \rho p_Z}
   -i \epsilon_{\mu \nu \rho q_Z}) .
\label{vector}
\end{equation}
It is to be noted that the coupling to the vector field
actually contains only two parameters and is therefore simpler
than to the scalar.

Using the generalised couplings given above we calculated the matrix
elements for $qQ\to q'Q' H $ where the primed and unprimed quarks
belong to the same $SU(2)$ doublet. In combination with the matrix
elements for $H\to WW\to \ell\nu\,\ell\nu$ given in \cite{Buszello:2002uu}
the full matrix element for $qQ \to q'Q' \ell\nu\,\ell\nu$ can be
obtained. Using this result an event generator was written to study
the effects of the various cuts. 
The QCD background has been simulated using
Pythia\cite{Sjostrand:2003wg} 
while we used Madgraph/Madevent\cite{Maltoni:2002qb} for the electroweak
background.

%% file: chapter3.tex
\section{Reconstruction and analysis}
The ATLAS collaboration has demonstrated, that the 
H $\rightarrow$ WW $\rightarrow \ell\nu\,\ell\nu$ signal can be reconstructed 
very well above a very small background using only 30 fb$^{-1}$
\cite{Asai:2004ws}. This makes this channel a very promising candidate for
the discovery of the Higgs boson.
We will use the cuts described there and  normalise our samples to the 
number of signal and background events found in that very detailed study.
We will discuss the effect of some of the cuts on the results we get, 
and vary them to show how a bias in the measured quantities can be reduced 
in exchange for a higher background level.

We will give a short summary of the main features of the signal selection, 
but not repeat the cuts in all detail (please refer to \cite{Asai:2004ws}).
We require two leptons with high transverse momenta (P$_T^1 >$ 20 GeV and P$_T^2 >$15 GeV) to ensure
that the events can be reliably triggered on.
The main part of the reconstruction consists of the selection of two 
jets in the forward direction (the tag jets) with a large separation in rapidity,
combined with a veto on hard jets in the rapidity region between the tag jets.  
This exploits the fact that 
the signal process lacks colour flow in the t-channel, unlike the most 
prominent background - namely $t\bar{t}$ production. A moderate cut 
on the  dijet mass is imposed. The jets are used to balance the momentum 
of the decay products of the Higgs, so that a cut on the transverse momentum 
of this system consisting of the two jet, the leptons and the missing 
transverse momentum can be applied. 
We will use the angle between the two jets as one of the 
variables, that allows us to determine if the spin/CP state of the particle
found is consistent with that of the SM Higgs boson.

The other major aspect of the ATLAS analysis deals with correlations of
the decay leptons. Based on a spin/CP hypothesis of 0$^{+}$ for the
Higgs boson, the decay products will have certain angular correlations,
that differ from those of the background processes.  Cuts on the angle
$\phi$ between the two leptons in the transverse plane, on the
separation in rapidity $\eta$ and $\phi$, the cosine of the polar
opening angle and the mass of the di-lepton system are applied in order
to maximise the signal to background ratio. Obviously, these cuts
interfere with the determination of spin and CP. We will study the
effect of these cuts on the
significance of the spin and CP measurement.

The potential background from Z$\rightarrow \tau\tau$ can be suppressed by 
trying to reconstruct the $\tau$ leptons in a collinear approximation using the missing 
transverse momentum, and imposing a lower limit on the transverse mass of the 
di-lepton and neutrino system. While the Z$\rightarrow \tau\tau$ background 
can be effectively suppressed by this, the cuts used have a negligible 
effect on the quantities used in this analysis.\\
For the further discussion we define the following subsets of cuts:\\
The ``tagging cuts'' are those, that define the tag jets and select the signal using these jets:
\begin{itemize}
\item Two leptons with transverse momenta $P_T > 15$ GeV and one lepton with $P_T > 20$ GeV and rapidity $|\eta| < 2.5$
\item Two jets with $P_T^1 > 40$ GeV  and $P_T^2 > 20$ GeV with a separation in rapidity $\delta \eta > 3.8$  and the two leptons lie within this gap.
\item The dijet mass $M_{jj} >$ 550 GeV 
\end{itemize}
The ``lepton cuts'' further suppress the background and can  bias the spin/CP measurements:
\begin{itemize}
\item The angle between the two leptons in the transverse plane $\Delta\phi_{\ell\ell} \le 1.05$
\item Separation in the $\eta$-$\phi$-plane $\Delta R_{\ell\ell} \le 1.8$
\item Cosine of the polar opening angle $\cos\theta_{\ell\ell} \ge 0.2$
\item Di-lepton mass $M_{\ell\ell} < $ 85 GeV
\item Transverse momentum $P_T(\ell_{1,2}) < $ 120 GeV  
\end{itemize}
\begin{figure}[htbp]
  \includegraphics[width=7cm]{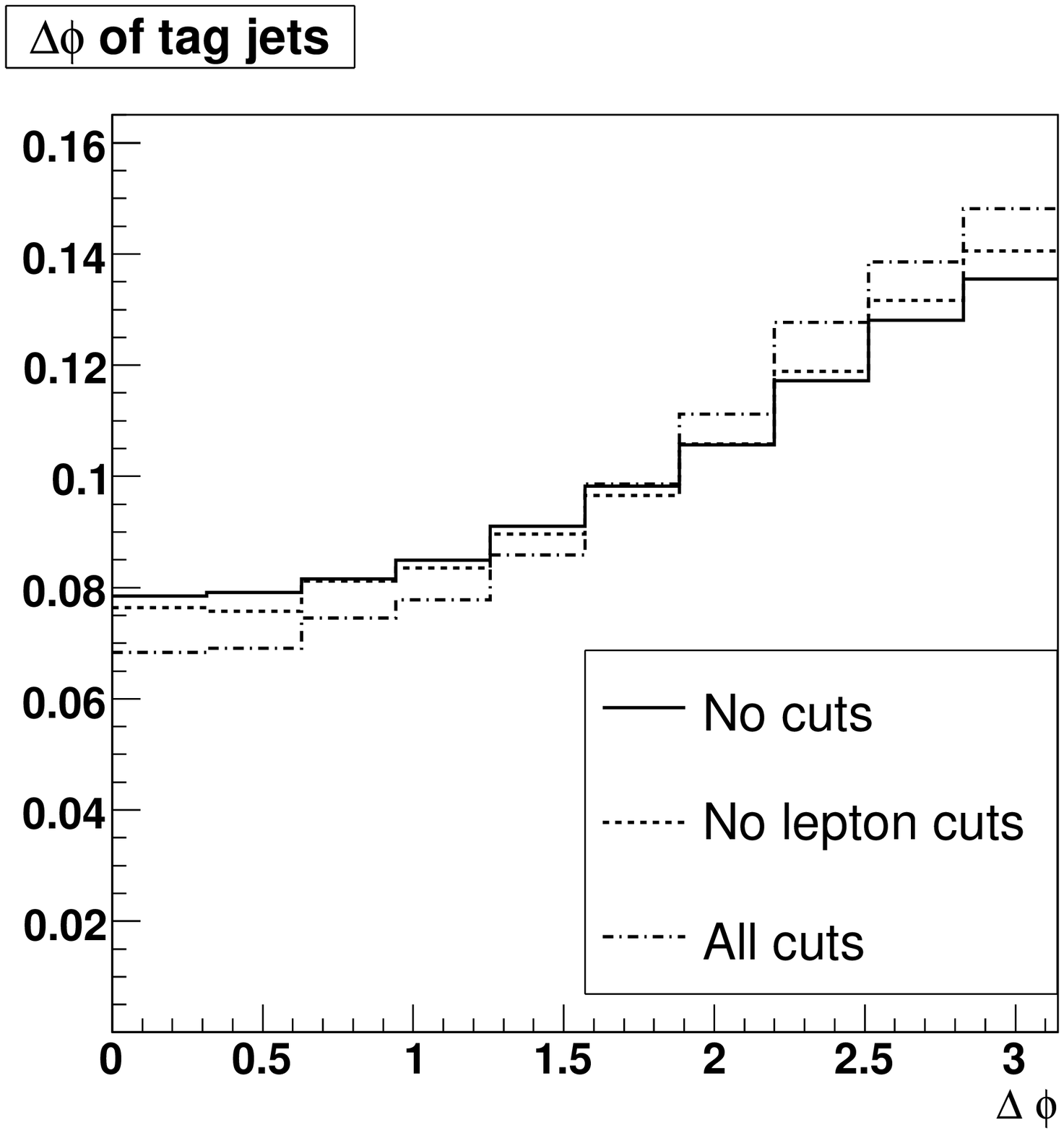}
  \includegraphics[width=7cm]{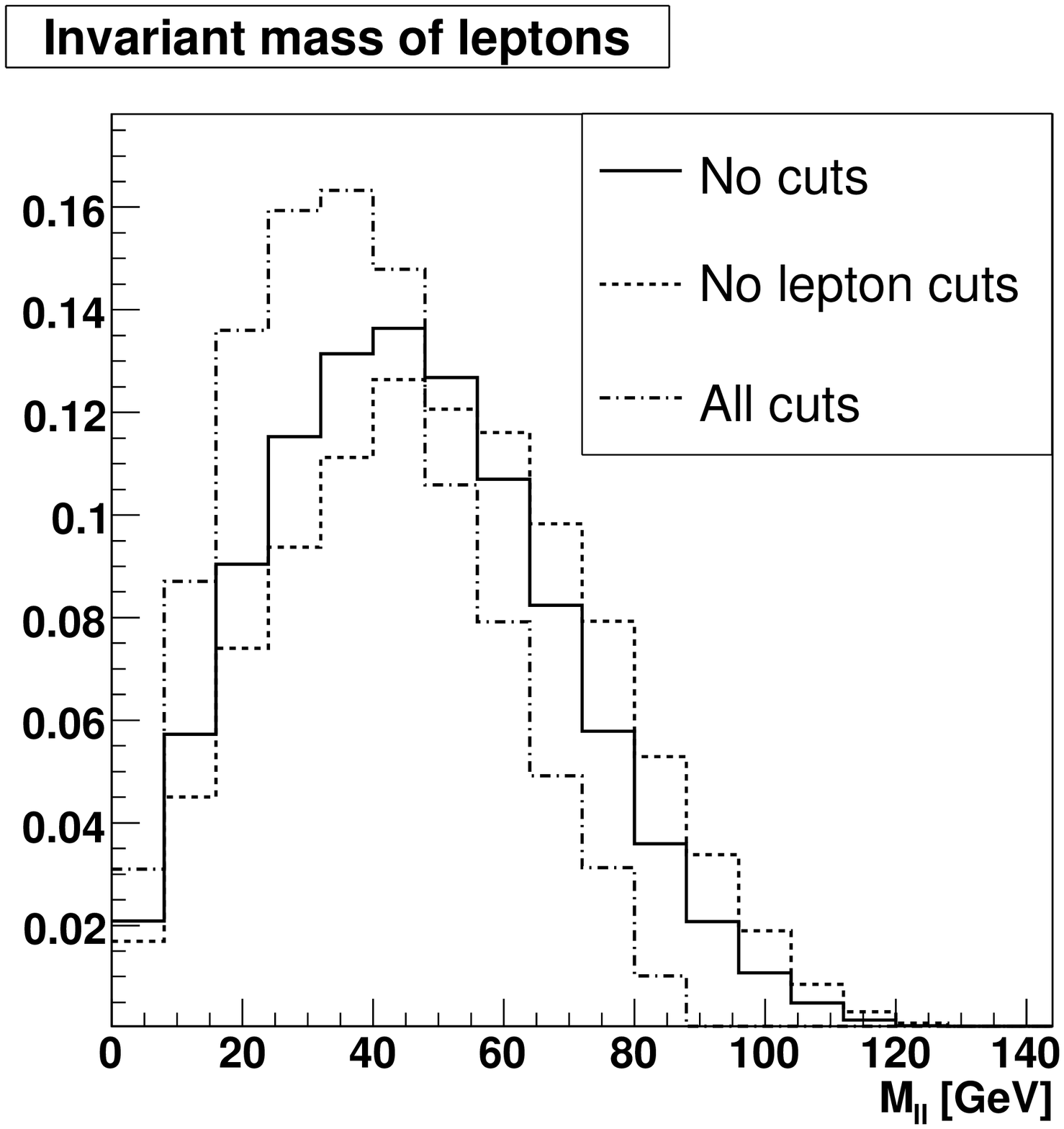}
  \caption{The effect of the cuts on the jet angle (left) and di-lepton mass (right) 
    distributions of the SM Higgs assuming a Higgs mass of 150 GeV.
    The solid lines reflect the distribution at tree level without any cuts applied. 
    The dashed line shows the distributions with only the tagging cuts applied
    and the dash-dotted one with all cuts.}
  \label{cuteffect}
\end{figure}
\subsection{Tag jet angle}
\begin{figure}
  \includegraphics[width=7cm]{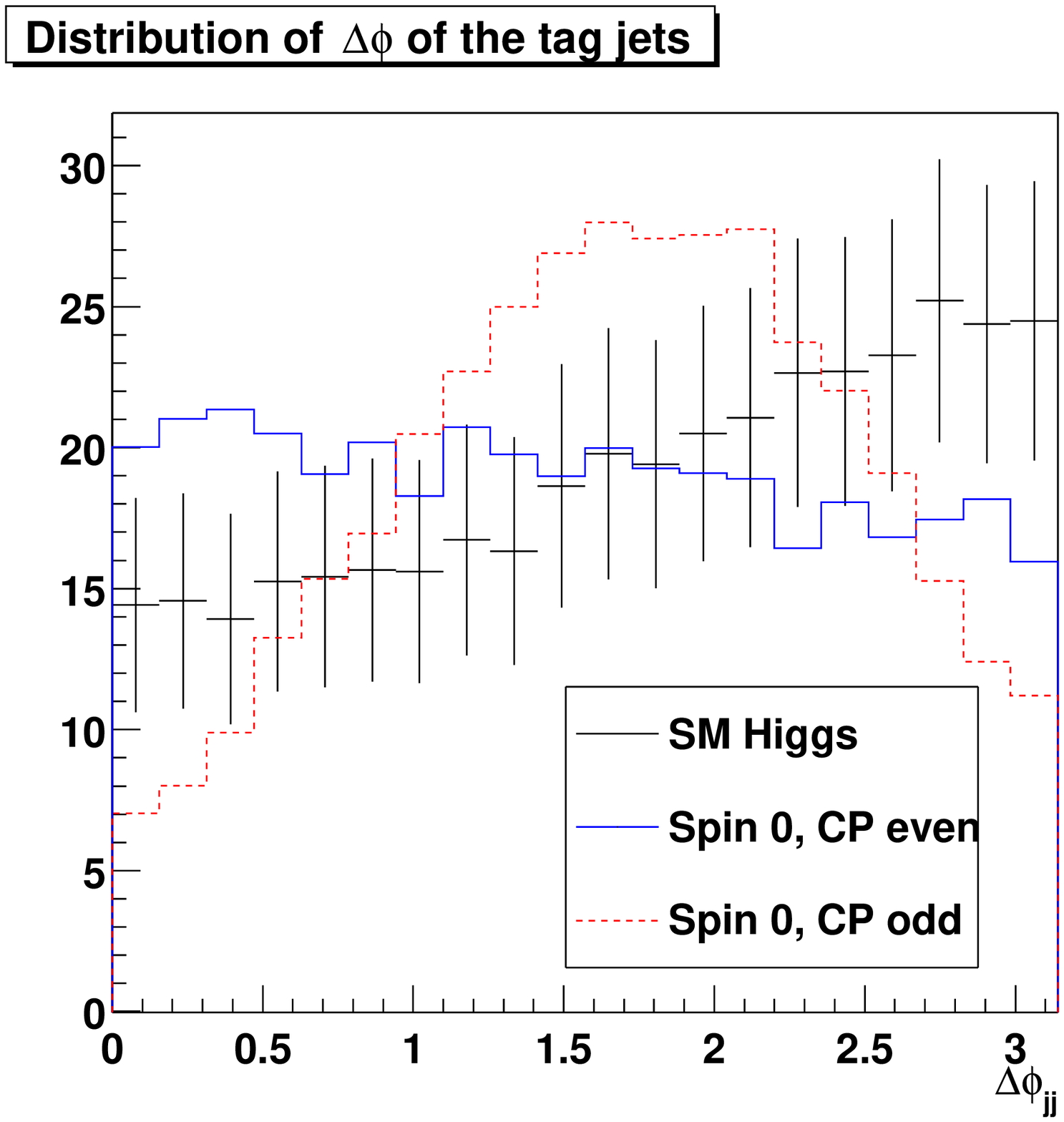}
  \includegraphics[width=7cm]{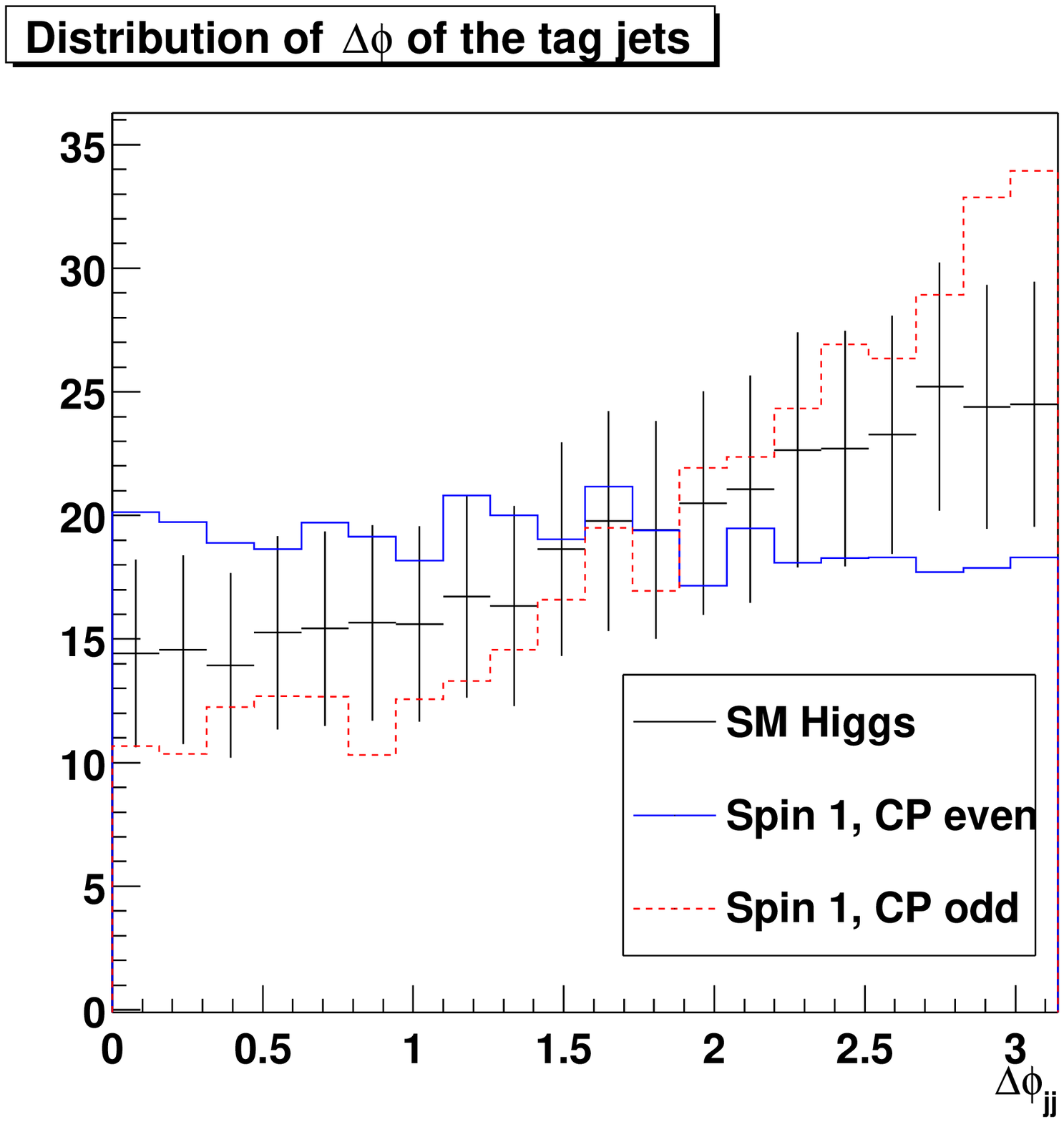}
  \caption{The distributions of the tag jet angle for the non SM couplings with a 
    Higgs mass of 150 GeV. The points with error-bars show the SM distribution.
    The errorbars indicate the expected statistical error with an integrated luminosity 
    of 30 fb$^{-1}$ neglecting the background contribution.}
  \label{jetangles}
\end{figure}
While the rapidity gap between the forward jets in WBF is due to lack of
colour flow, the angle between the jets highly depends on the coupling
structure of the Higgs to the vector bosons. This can be exploited to
determine the structure of the coupling by comparing the measured
distribution of the selected tag jets to the distributions expected for
non SM couplings. Some of the cuts used to select the signal process
have the potential to bias the distribution.  A lower limit on the dijet
mass should favour higher angles between the jets, and the cuts on the
decay particles could have different effects.  Figure \ref{cuteffect}
illustrates the effect of different sets of cuts on the tag jet angle for
a Higgs boson mass of 150 GeV.  The solid line shows the distribution on parton
level with no cuts applied. When all but the lepton cuts are applied the
distribution becomes slightly steeper (dashed) and when finally all cuts
are applied (dash-dotted) it steepens even further. But overall the
effects are rather small, and do not diminish the power of this variable
to discriminate between different spin/CP states.  Figure 
\ref{jetangles} shows the distribution of the tag jet angle for various
spin/CP states and a Higgs mass of 150 GeV. The error-bars indicate the
expected number of events for a SM Higgs and the expected statistical
error when using the results of \cite{Asai:2004ws}. The dashed and solid lines 
show the expected distributions of the other spin 0 (left) and spin
1 (right) hypothesis. All histograms are normalised to the expected number of
events in the SM as any significant deviation would rule out a SM Higgs
directly. All non SM couplings show significant deviations
from the SM case, especially the spin 0, CP odd case.

\begin{figure}
\centering
\includegraphics[width=15cm]{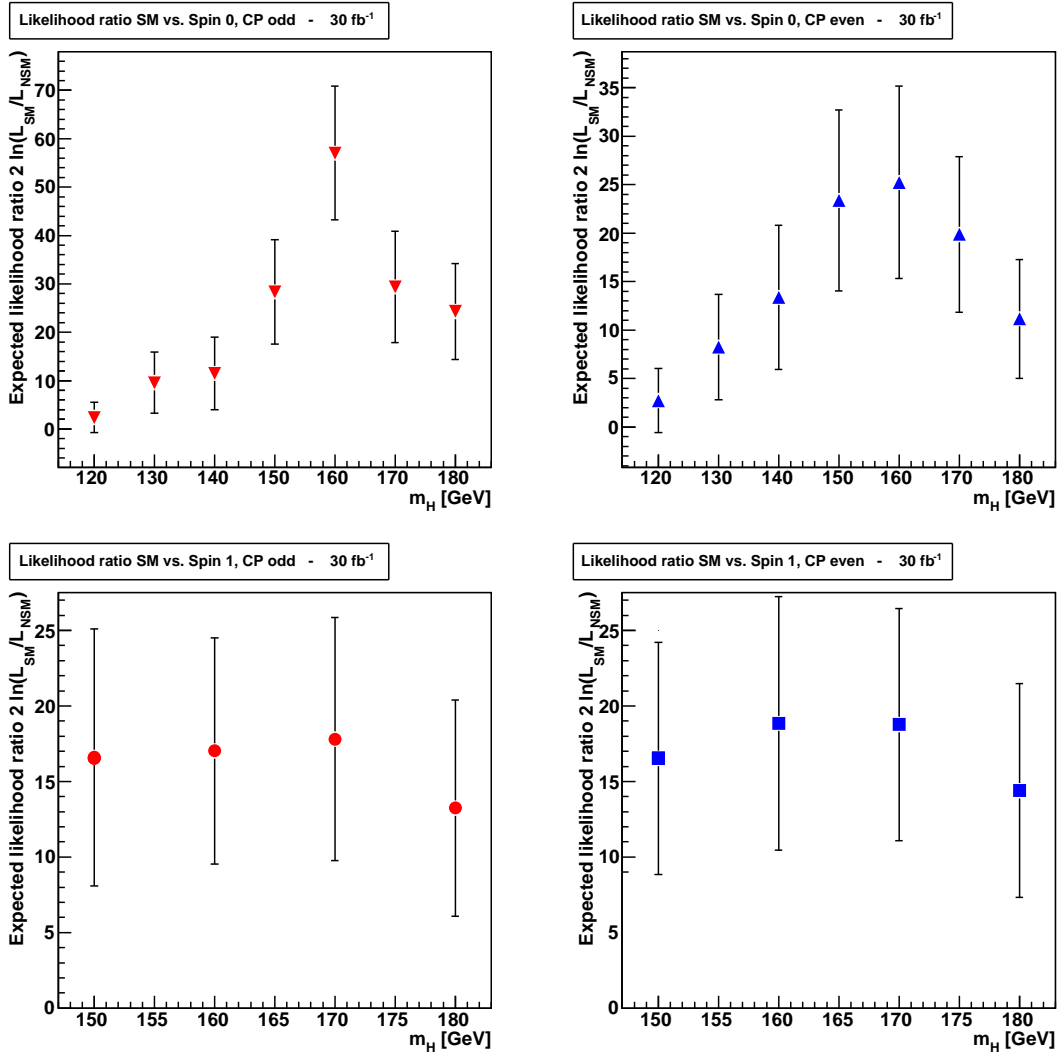}
\caption{Exclusion significance from jet angles using 30 fb$^{-1}$. The points show the mean log likelihood ratio 
$2\mathrm{ln} (L_{SM}/L_{NSM})$. The error-bars reflect the RMS of the distribution of the likelihood ratio. 
2/3 of a large number of experiments would yield a result within this range.}
\label{likhood}
\end{figure}

To evaluate the feasibility of the measurement, we use simulated samples
of SM Higgs signal events and samples of background events, where the
number of events in these samples is taken from \cite{Asai:2004ws}. We
then compare the tag jet angle distribution of these samples with SM and
non SM distributions derived from larger MC samples.  All samples have
passed the same complete list of cuts, and thus the potential bias
introduced by those cuts is modeled correctly. We use a binned
likelihood to evaluate how significant the deviation from the non SM
models is.  For large event numbers the ratio $2\mathrm{ln}
(L_{SM}/L_{NSM})$ corresponds to the standard deviation of a $\chi^2$
test.  In figure \ref{likhood} we plot the mean likelihood ratio for a
large number of MC experiments and the RMS of the distributions of the
likelihood ratio.

\subsection{Di-lepton mass}
\begin{figure}
  \includegraphics[width=7cm]{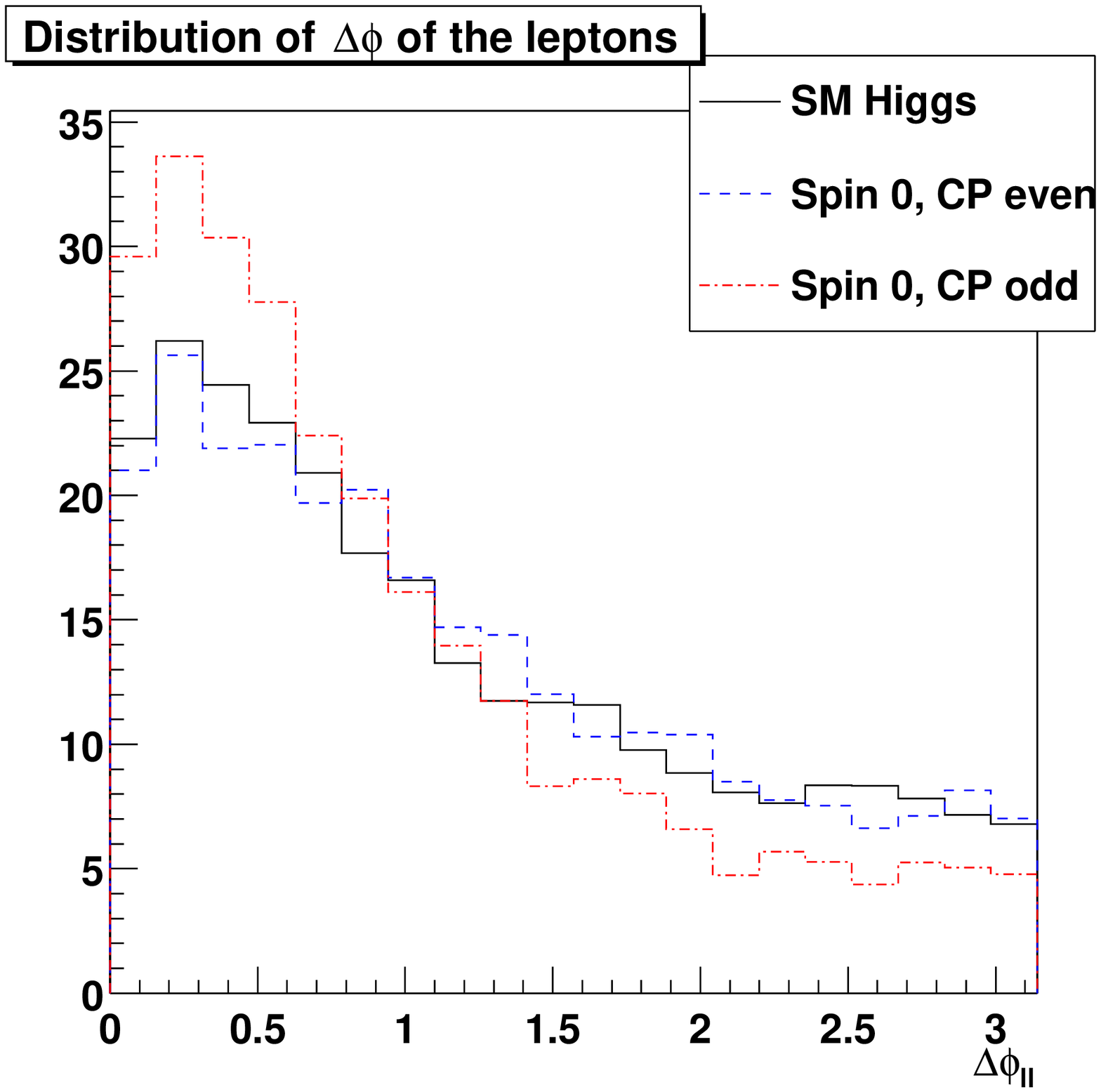}
  \includegraphics[width=7cm]{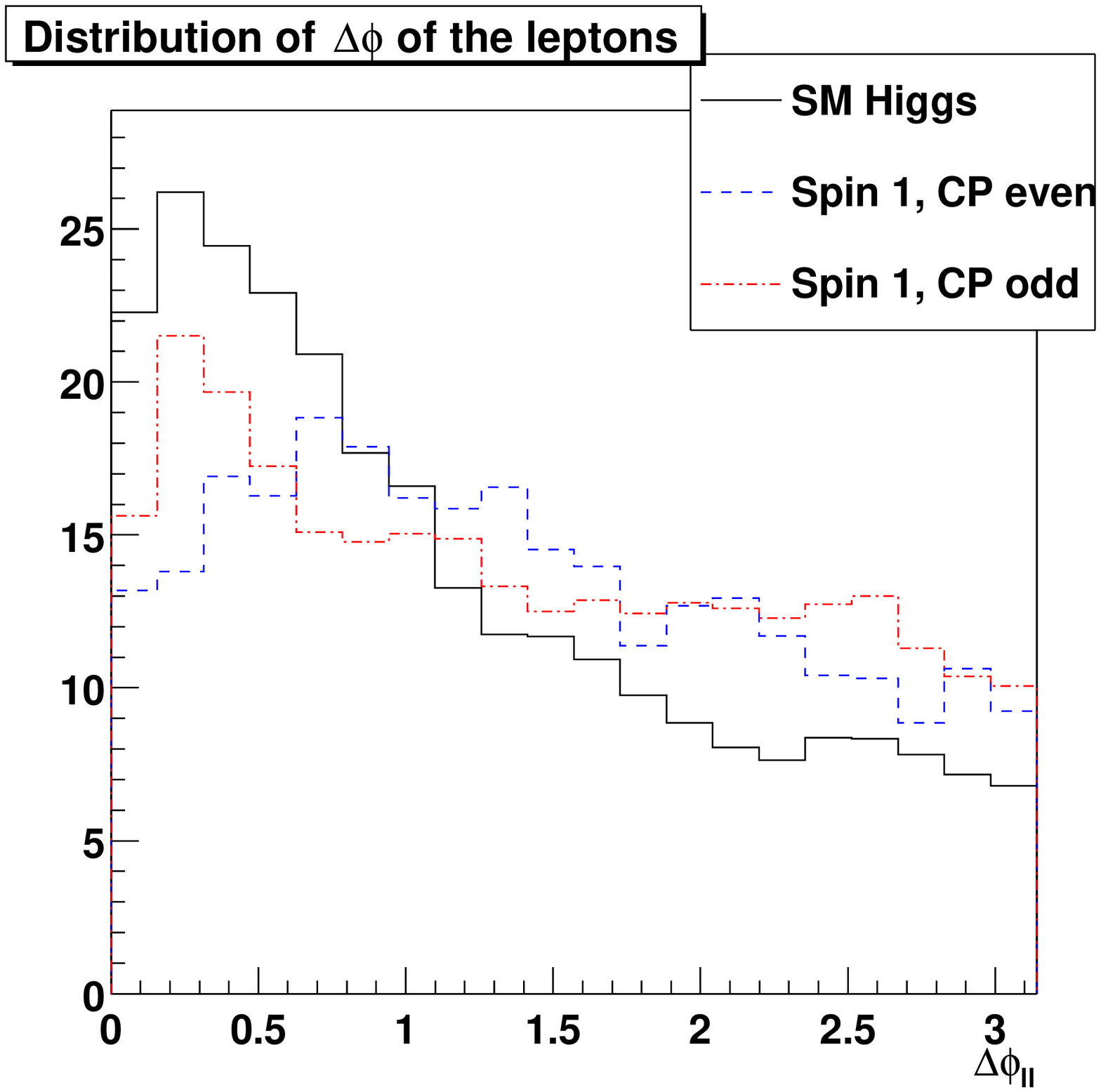}
\caption{The distribution of the angle between the leptons in the transverse plane 
for the SM and non SM couplings and a Higgs mass of 150 GeV. All distributions peak around 0. ISR will have a direct effect on
these distributions.}
\label{leptonangles}
\end{figure}

\begin{figure}
  \includegraphics[width=7cm]{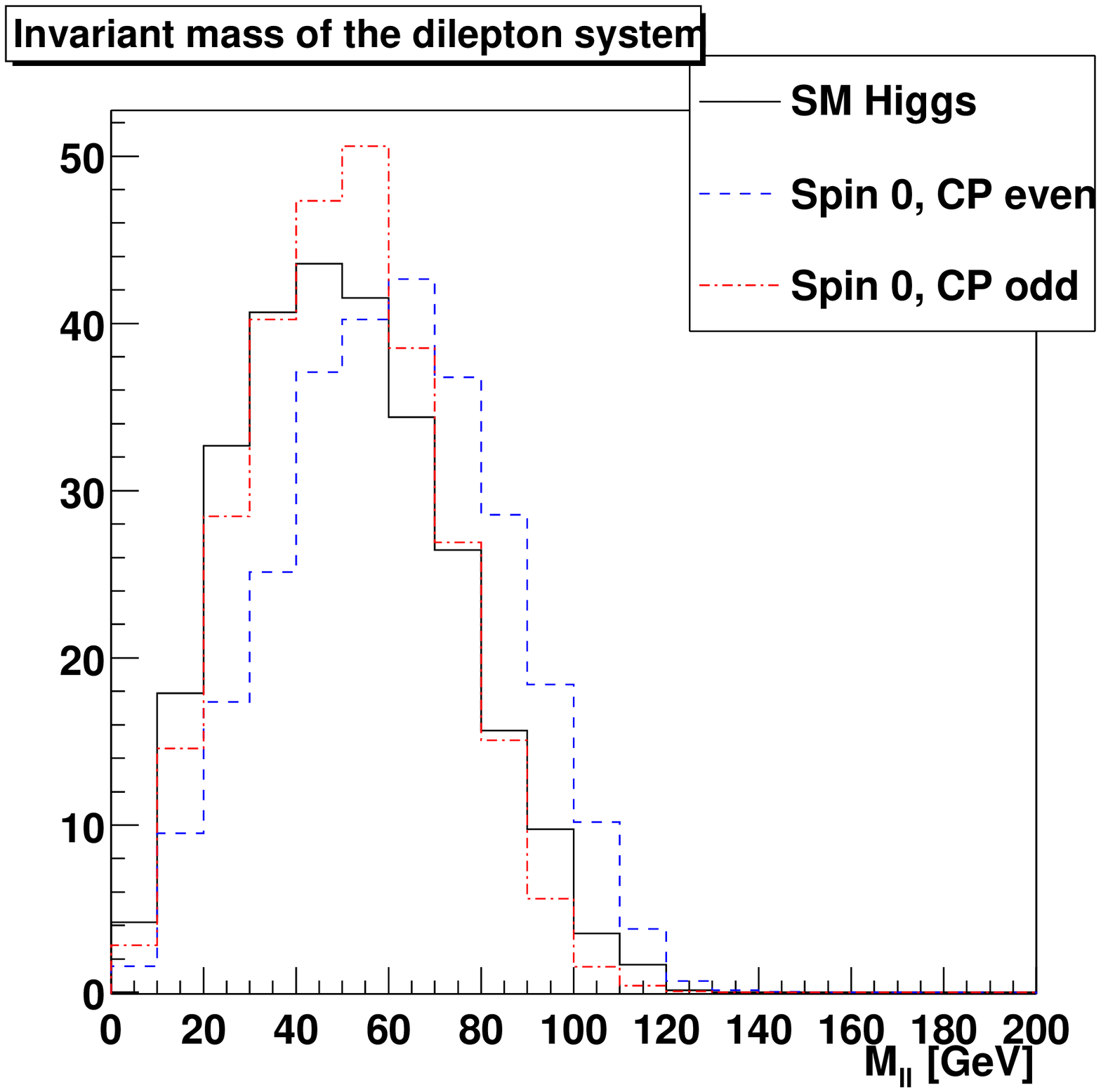}
  \includegraphics[width=7cm]{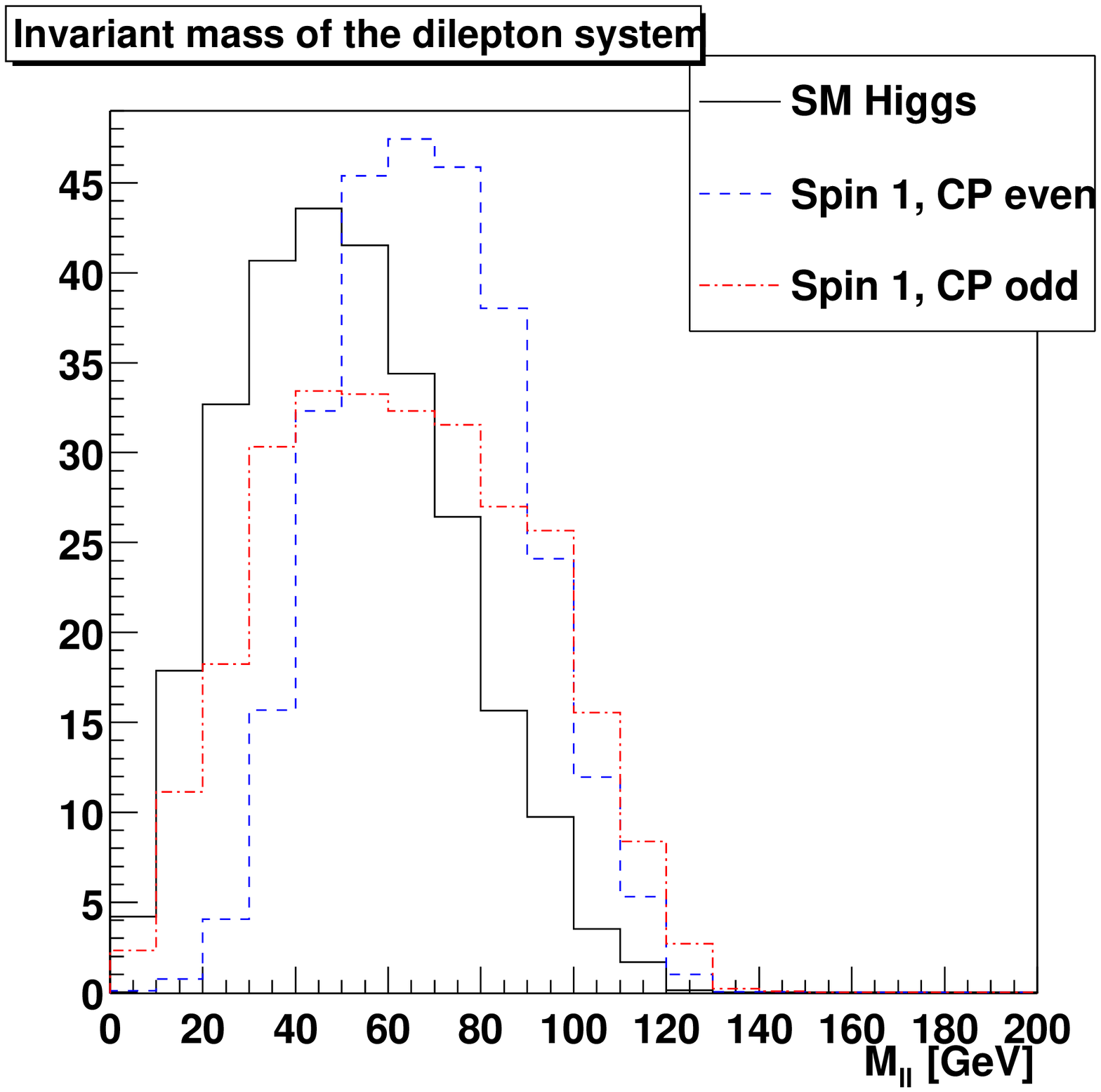}
\caption{The distribution of the di-lepton mass for SM and non SM couplings and a Higgs mass of 150 GeV. 
The mass is highly correlated to the lepton angle but invariant, so that ISR will not affect the distribution.}
\label{dileptonmass}
\end{figure}

It has been suggested \cite{Asai:2004ws} that the distribution of the 
angle between the leptons in the transverse plane could be used to 
verify the SM spin/CP hypothesis. Unfortunately, the SM and the non SM 
distributions all peak at small angles (see figure \ref{leptonangles}). 
Furthermore, the angle is obviously not 
Lorentz invariant, so that an initial state boost would change the distributions.
Considering potential problems in modeling initial state radiation (ISR) at the LHC, 
we would forfeit the 
main advantage of using the very clean and well measurable leptonic decay products.
We avoid such complications by using the invariant di-lepton mass which is correlated 
to the angular distributions, and peaks at different values for the various scenarios
(figure \ref{dileptonmass}). Probably the most straight forward variable to use 
is the mean value of the invariant mass distribution. Again, we will determine 
the feasibility using  a number of MC samples and determine the mean and RMS of the 
distribution of the mean di-lepton mass.

The most problematic aspect of this analysis is the obvious bias introduced by 
the lepton cuts (see figure \ref{cuteffect}). Since larger angles are correlated with a larger 
invariant mass of the di-lepton system, the cuts on the angle in the transverse plane and the 
cut on the separation in the R-$\phi$-plane bias the distribution significantly and 
shift the mean towards smaller values for all couplings. 
The distinguishing power of this variable is thereby decreased.

We list the expected mean 
for the various couplings and different sets of cuts in table 
\ref{tabcuts} for a Higgs mass of 160 GeV. For clarity we neglect any background  
for the numbers in this table. They would introduce varying offsets for the different sets of
cuts. Of course, these effects are included when determining the exclusion 
significances, where they have a minor
effect, as they mostly cancel out in the mean mass differences.

We define a set of relaxed lepton cuts as follows (numbers in brackets show the standard values 
as determined in \cite{Asai:2004ws}): The cut on the angle of the leptons in the 
transverse plane is changed to  $\Delta\phi_{\ell\ell} \le 1.9$ (was 1.05), 
the separation in the $\eta$-$\phi$-plane to $\Delta R_{\ell\ell} \le 2.5$ (was 1.8),
the cosine of the polar opening angle to $\cos\theta_{\ell\ell} \ge 0.0$ (was 0.2).
The signal region is defined by the di-lepton mass  $M_{\ell\ell} < $ 100 GeV (was 85 GeV) 
and the transverse mass $m_{\ell\ell\nu} < $180 GeV (was 175 GeV).
The cut on the upper bound of the lepton $P_T$ remains unchanged.
The more extreme case where the cuts on $\Delta\phi_{\ell\ell}$, $\Delta R_{\ell\ell}$
and  $\cos\theta_{\ell\ell}$ are completely dropped and the signal region is widened even further 
to   $M_{\ell\ell} < $ 120 GeV and $m_{\ell\ell\nu} < $ 200 GeV will be called ``no lepton cuts''.
Loosening the cuts on these variables obviously leeds to a 
rise in the expected background. 
In table \ref{tabcuts} we list the increase of the background relative to the 
expected background given in \cite{Asai:2004ws}. As we do not use 
any detector simulation, we are not able to reproduce the absolute number of events expected,
but the relative numbers given here are reliable, since they depend mostly on 
cuts on  very well measured quantities related to the high $P_T$ leptons.
The largest contribution to the background comes from top pair
production and the electroweak  WW processes.
The QCD WW production is well enough suppressed by cuts on the tag jets, so that even with the rather high 
relative efficiency from the relaxed cuts it still doesn't contribute much to the background. 
Taking all three backgrounds and their respective fractions of the total background (from \cite{Asai:2004ws}) 
into account we can estimate the increase of the background to be
from 2.5 for the relaxed cuts to 4.5 with no lepton cuts. This and the rise in signal are compatible with what 
can be seen in the plots in \cite{Asai:2004ws}.
\begin{table}
  \begin{tabular*}{\textwidth}{l | c c c c c || c c c c} 
       \hline \multicolumn{1}{c|}{} &  \multicolumn{5}{c||}{mean $M_{\ell\ell}$/GeV} &  \multicolumn{4}{c}{relative efficiencies}\\\hline
    cuts & SM &  0$^+$ &  0$^-$        & 1$^+$ & 1$^-$ &  Signal& tt& WW(qcd)& WW(ew)\\ \hline \hline
    Standard  &36.7 &  43.3 & 40.7 &53.3 &39.4 &1 &1 & 1 & 1\\ \hline 
    Relaxed & 45.3 &56.2&  48.5&65.3 & 52.4 & 1.92 & 2.47 & 2.80 & 1.78\\ \hline 
    No lepton & 50.2&62.2&50.7& 70.2&63.9 & 2.90 & 4.48 & 6.76 & 3.80\\ \hline 
  \end{tabular*}
  \caption{The effect of different sets of cuts on signal and background, and on the expected mean of the di-lepton mass for 
 a Higgs boson of 160 GeV.}
  \label{tabcuts}
\end{table}

\begin{figure}
  \includegraphics[width=7cm]{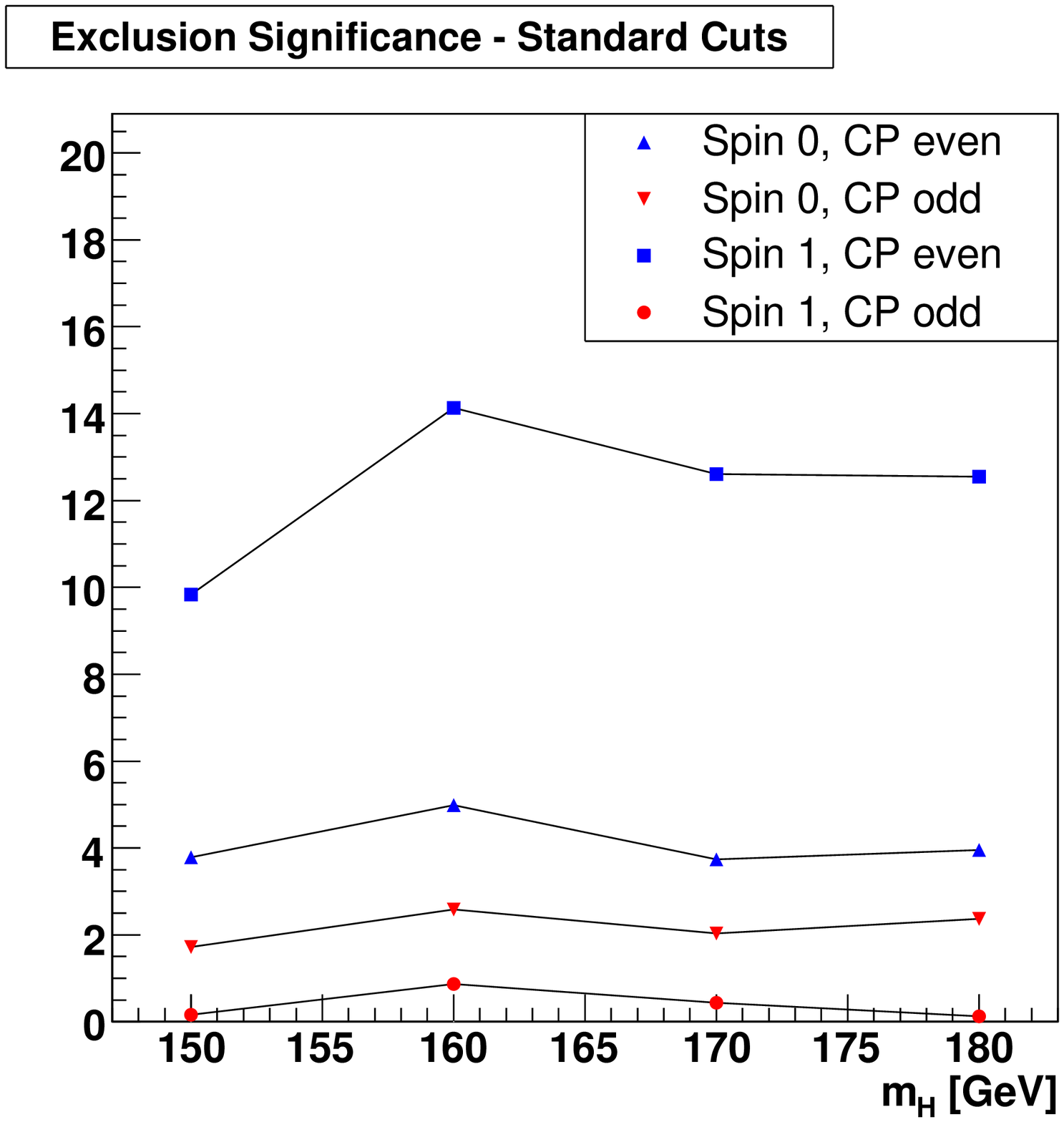}
  \includegraphics[width=7cm]{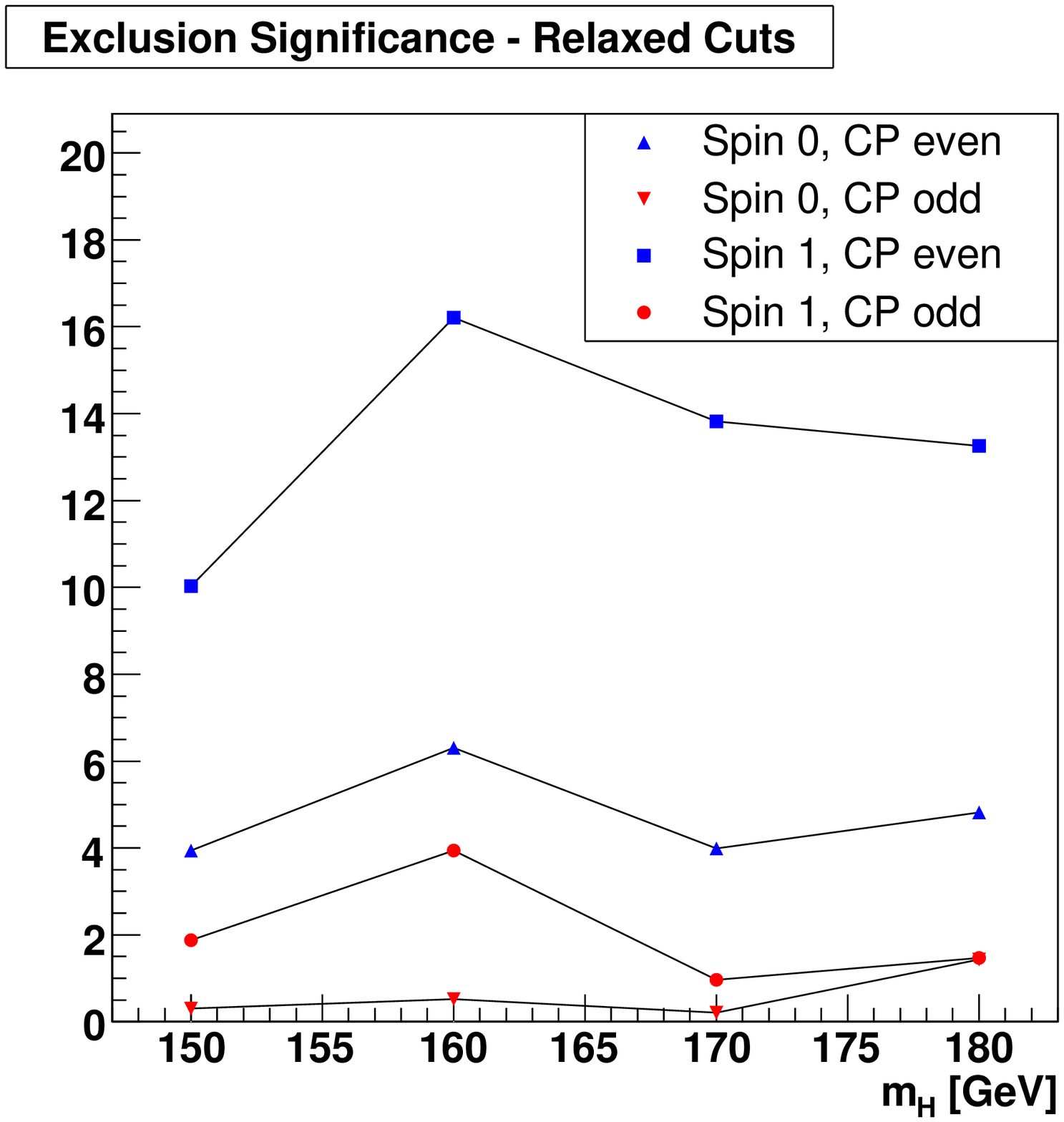}
 \includegraphics[width=7cm]{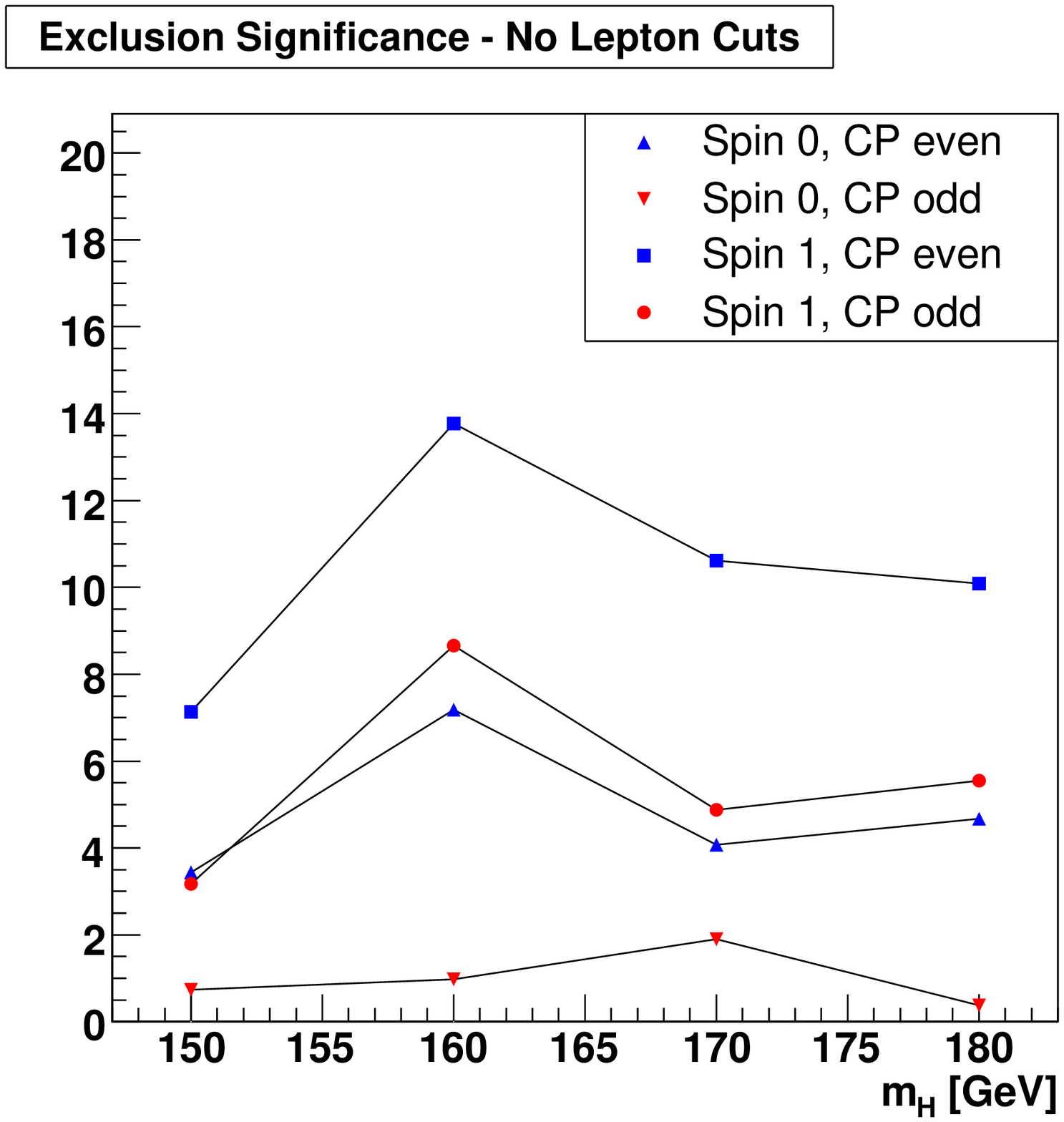}
\caption{The expected exclusion significance of non SM couplings, using
three different sets of cuts. The standard set of cuts as defined in
\cite{Asai:2004ws} (upper left),
the relaxed lepton cuts (upper right) and no lepton cuts (bottom).}

\label{dilepres}
\end{figure}

While the lepton cuts reduce the background, they also reduce the number of signal events and the 
differences in the di-lepton mass. We explore the effect of the three different cuts sets, taking 
into account the effects on the distributions and the change in number of signal and background 
events. We use samples with the expected number of signal and background events and calculate the
mean of the di-lepton mass. Doing this for many samples, we then use the mean of those means 
as the expected value, and the RMS of the means as the expected error. In figure \ref{dilepres}, we  plot 
the difference between SM and non-SM values divided by the expected error. From these we can see, that the 
spin 1, CP even case can be ruled out using the di-lepton mass with any of the sets of cuts,
while the Spin 1 CP odd case becomes too similar to the SM case, when applying any lepton cuts. 
However, it can be ruled out for some Higgs masses when the lepton cuts are completely dropped.

